\documentclass[twoside,psfig]{article}
\usepackage{graphicx}
\catcode`\@=11
\long\def\@makefntext#1{
\protect\noindent \hbox to 3.2pt {\hskip-.9pt
$^{{\eightrm\@thefnmark}}$\hfil}#1\hfill}       

\def\@makefnmark{\hbox to 0pt{$^{\@thefnmark}$\hss}}    

\def\ps@myheadings{\let\@mkboth\@gobbletwo
\def\@oddhead{\hbox{}
\rightmark\hfil\eightrm\thepage}
\def\@oddfoot{}\def\@evenhead{\eightrm\thepage\hfil
\leftmark\hbox{}}\def\@evenfoot{}
\def\sectionmark##1{}\def\subsectionmark##1{}}

\oddsidemargin=\evensidemargin
\newcounter{sectionc}\newcounter{subsectionc}\newcounter{subsubsectionc}
\renewcommand{\section}[1] {\vspace{12pt}\addtocounter{sectionc}{1}
\setcounter{subsectionc}{0}\setcounter{subsubsectionc}{0}\noindent
    {\tenbf\thesectionc. #1}\par\vspace{5pt}}
\renewcommand{\subsection}[1] {\vspace{12pt}\addtocounter{subsectionc}{1}
    \setcounter{subsubsectionc}{0}\noindent
    {\bf\thesectionc.\thesubsectionc. {\kern1pt \bfit #1}}\par\vspace{5pt}}
\renewcommand{\subsubsection}[1] {\vspace{12pt}\addtocounter{subsubsectionc}{1}
    \noindent{\tenrm\thesectionc.\thesubsectionc.\thesubsubsectionc.
    {\kern1pt \tenit #1}}\par\vspace{5pt}}
\newcommand{\nonumsection}[1] {\vspace{12pt}\noindent{\tenbf #1}
    \par\vspace{5pt}}

\newcounter{appendixc}
\newcounter{subappendixc}[appendixc]
\newcounter{subsubappendixc}[subappendixc]
\renewcommand{\thesubappendixc}{\Alph{appendixc}.\arabic{subappendixc}}
\renewcommand{\thesubsubappendixc}
    {\Alph{appendixc}.\arabic{subappendixc}.\arabic{subsubappendixc}}

\renewcommand{\appendix}[1] {\vspace{12pt}
        \refstepcounter{appendixc}
        \setcounter{figure}{0}
        \setcounter{table}{0}
        \setcounter{lemma}{0}
        \setcounter{theorem}{0}
        \setcounter{corollary}{0}
        \setcounter{definition}{0}
        \setcounter{equation}{0}
        \renewcommand{\thefigure}{\Alph{appendixc}.\arabic{figure}}
        \renewcommand{\thetable}{\Alph{appendixc}.\arabic{table}}
        \renewcommand{\theappendixc}{\Alph{appendixc}}
        \renewcommand{\thelemma}{\Alph{appendixc}.\arabic{lemma}}
        \renewcommand{\thetheorem}{\Alph{appendixc}.\arabic{theorem}}
        \renewcommand{\thedefinition}{\Alph{appendixc}.\arabic{definition}}
        \renewcommand{\thecorollary}{\Alph{appendixc}.\arabic{corollary}}
        \noindent{\tenbf Appendix \theappendixc #1}\par\vspace{5pt}}
\newcommand{\subappendix}[1] {\vspace{12pt}
        \refstepcounter{subappendixc}
        \noindent{\bf Appendix \thesubappendixc. {\kern1pt \bfit #1}}
    \par\vspace{5pt}}
\newcommand{\subsubappendix}[1] {\vspace{12pt}
        \refstepcounter{subsubappendixc}
        \noindent{\rm Appendix \thesubsubappendixc. {\kern1pt \tenit #1}}
    \par\vspace{5pt}}

\newcommand{\textlineskip}{\baselineskip=13pt}
\newcommand{\smalllineskip}{\baselineskip=10pt}

\def\eightcirc{
\begin{picture}(0,0)
\put(4.4,1.8){\circle{6.5}}
\end{picture}}
\def\eightcopyright{\eightcirc\kern2.7pt\hbox{\eightrm c}}

\newcommand{\copyrightheading}[1]
    {\vspace*{-2.5cm}\smalllineskip{\flushleft
    {\footnotesize International Journal of Modern Physics C #1}\\
    {\footnotesize $\eightcopyright$\, World Scientific Publishing
     Company}\\
     }}

\newcommand{\publisher}[2]{{\begin{center}\footnotesize\smalllineskip
    Received #1\\
    Revised #2
    \end{center}
    }}

\def\abstracts#1#2#3{{
    \centering{\begin{minipage}{4.5in}\footnotesize\baselineskip=10pt
    \parindent=0pt #1\par
    \parindent=15pt #2\par
    \parindent=15pt #3
    \end{minipage}}\par}}

\def\keywords#1{{
    \centering{\begin{minipage}{4.5in}\footnotesize\baselineskip=10pt
    {\footnotesize\it Keywords}\/: #1
    \end{minipage}}\par}}

\newcommand{\bibit}{\nineit}
\newcommand{\bibbf}{\ninebf}
\renewenvironment{thebibliography}[1]
        {\frenchspacing
     \ninerm\baselineskip=11pt
         \begin{list}{\arabic{enumi}.}
        {\usecounter{enumi}\setlength{\parsep}{0pt}
     \setlength{\leftmargin 12.7pt}{\rightmargin 0pt} 
         \setlength{\itemsep}{0pt} \settowidth
    {\labelwidth}{#1.}\sloppy}}{\end{list}}

\newcounter{itemlistc}
\newcounter{romanlistc}
\newcounter{alphlistc}
\newcounter{arabiclistc}

\newcommand{\fcaption}[1]{
        \refstepcounter{figure}
        \setbox\@tempboxa = \hbox{\footnotesize Fig.~\thefigure. #1}
        \ifdim \wd\@tempboxa > 5in
           {\begin{center}
        \parbox{5in}{\footnotesize\smalllineskip Fig.~\thefigure. #1}
            \end{center}}
        \else
             {\begin{center}
             {\footnotesize Fig.~\thefigure. #1}
              \end{center}}
        \fi}

\newcommand{\tcaption}[1]{
        \refstepcounter{table}
        \setbox\@tempboxa = \hbox{\footnotesize Table~\thetable. #1}
        \ifdim \wd\@tempboxa > 5in
           {\begin{center}
        \parbox{5in}{\footnotesize\smalllineskip Table~\thetable. #1}
            \end{center}}
        \else
             {\begin{center}
             {\footnotesize Table~\thetable. #1}
              \end{center}}
        \fi}

\def\@citex[#1]#2{\if@filesw\immediate\write\@auxout
    {\string\citation{#2}}\fi
\def\@citea{}\@cite{\@for\@citeb:=#2\do
    {\@citea\def\@citea{,}\@ifundefined
    {b@\@citeb}{{\bf ?}\@warning
    {Citation `\@citeb' on page \thepage \space undefined}}
    {\csname b@\@citeb\endcsname}}}{#1}}

\newif\if@cghi
\def\cite{\@cghitrue\@ifnextchar [{\@tempswatrue
    \@citex}{\@tempswafalse\@citex[]}}
\def\citelow{\@cghifalse\@ifnextchar [{\@tempswatrue
    \@citex}{\@tempswafalse\@citex[]}}
\def\@cite#1#2{{$\null^{#1}$\if@tempswa\typeout
    {IJCGA warning: optional citation argument
    ignored: `#2'} \fi}}

\def\pmb#1{\setbox0=\hbox{#1}
    \kern-.025em\copy0\kern-\wd0
    \kern.05em\copy0\kern-\wd0
    \kern-.025em\raise.0433em\box0}

\def\fnt#1#2{\footnotetext{\kern-.3em
    {$^{\mbox{\scriptsize #1}}$}{#2}}}

\def\ps@myheadings{%
    \let\@oddfoot\@empty\let\@evenfoot\@empty
    \def\@evenhead{\slshape\leftmark\hfil}
    \def\@oddhead{\hfil{\slshape\rightmark}}
    \let\@mkboth\@gobbletwo
    \let\sectionmark\@gobble
    \let\subsectionmark\@gobble
    }
\font\tenrm=cmr10
\font\tenit=cmti10
\font\tenbf=cmbx10
\font\bfit=cmbxti10 at 10pt
\font\ninerm=cmr9
\font\nineit=cmti9
\font\ninebf=cmbx9
\font\eightrm=cmr8

\def\qed{\hbox{${\vcenter{\vbox{            
   \hrule height 0.4pt\hbox{\vrule width 0.4pt height 6pt
   \kern5pt\vrule width 0.4pt}\hrule height 0.4pt}}}$}}

\def\bsc{{\sc a\kern-6.4pt\sc a\kern-6.4pt\sc a}}   
\def\bflatex{\bf L\kern-.30em\raise.3ex\hbox{\bsc}\kern-.14em
T\kern-.1667em\lower.7ex\hbox{E}\kern-.125em X}

\begin{document}
\setlength{\textheight}{7.7truein}  

\thispagestyle{empty}

\markboth{\protect{\footnotesize\it Mingfeng He et
al.}}{\protect{\footnotesize\it A predator-prey model based on the
fully parallel Cellular Automata}}

\normalsize\textlineskip

\setcounter{page}{1}

\copyrightheading{}         

\vspace*{0.88truein}

\centerline{\bf A PREDATOR-PREY MODEL BASED ON THE}
\vspace*{0.035truein} \centerline{\bf FULLY PARALLEL CELLULAR
AUTOMATA } \vspace*{0.37truein} \centerline{\footnotesize Mingfeng
 He} \baselineskip=12pt \centerline{\footnotesize\it
Department of Applied Mathematics, Dalian University of
Technology,}\centerline{\footnotesize\it
\centerline{\footnotesize\it Center of innovative education and
practice of Dalian University of Technology,}} \baselineskip=10pt
\centerline{\footnotesize\it Dalian 116024, China}

\vspace*{15pt}          
\centerline{\footnotesize Hongbo Ruan} \baselineskip=12pt
\centerline{\footnotesize\it Department of Civil Engineering,
Dalian University of Technology} \baselineskip=10pt
\centerline{\footnotesize\it Dalian 116024, China}

\vspace*{15pt}          
\centerline{\footnotesize Changliang Yu} \baselineskip=12pt
\centerline{\footnotesize\it Department of Automation, Dalian
University of Technology} \baselineskip=10pt
\centerline{\footnotesize\it Dalian 116024, China}

\vspace*{0.225truein} \publisher{(received date)}{(revised date)}

\vspace*{0.25truein} \abstracts{We presented a predator-prey
lattice model containing moveable wolves and sheep, which are
characterized by Penna double bit strings. Sexual reproduction and
child-care strategies are considered. To implement this model in
an efficient way, we build a fully parallel Cellular Automata
based on a new definition of the neighborhood. We show the roles
played by the initial densities of the populations, the mutation
rate and the linear size of the lattice in the evolution of this
model.}{}{}

\vspace*{5pt} \keywords{Evolution; Parallel; Cellular automata;
Predator-prey model.}


\vspace*{1pt}\textlineskip  
\section{Introduction}     
\vspace*{-0.5pt} \noindent The problem of dynamical relations
between two or more interacting populations already has a long
history. It started with the classic papers by Volterra $^1$ and
Lotka $^2$ describing the fluctuations in the fish catch in the
Adriatic.    Since then various models have been introduced in
order to consider different aspects of natural life, including
motion, birth and death processes, evolution and extinction
$^{3,4,7}$. But the unrealistic unlimited growth of a population
has been found in the classic Lotka-Volterra models. In order to
include the effect of food and space restrictions and to keep the
population within the carrying capacity the population can
sustain, Verhulst factor is imposed in models of $^{5, 6,
16}$.

Recently, there are some lattice models discussed this problem via
computer simulation to try to bring the models more realistic
$^{7-11}$. They all used the standard square two-dimensional
lattice on which predators and prey move and reproduce following a
set of rules. The lattice models are dynamic computational systems
that are discrete in space, time and state and whose behavior is
specified by rules governing local relationships. They also have a
benefit in being visually informative of the progress of dynamic
events. Especially, Verhulst factor can be replaced in a natural
way.$^{24}$

The Monte Carlo (MC) method is used to simulate the predator-prey
system.$^{7-11}$ In one Monte Carlo step, a random sequence of
discrete events is generated. Although MC simulations provide
tractable rules for the predator-prey model, they are not suitable
for efficient parallelization due to the random selection of
lattice cells. However, there is another important approach to
simulate discrete events of predator-prey system on lattice, the
Cellular Automata (CA). This approach is fully parallel in the
sense that all the lattice cells can be updated simultaneously.
This has the advantages that fewer random numbers are required,
and the global updating is easier to implement in a parallel code
$^{21}$. A kinematic, asynchronous, stochastic cellular automata
was proposed to model water and solution phenomena encountered in
complex biological systems.$^{12}$ A stochastic cellular automata
is built to test sensitivity of model response under different
spatial and temporal sequences of events.$^{19}$ A probabilistic
cellular automata method was adopted to model a lattice
system.$^{20}$ It may be viewed as an automata network with a
mixed transition rule, at each time step the evolution results
from the application of the synchronous subrule followed by the
sequential one.

In this paper we presented a parallel cellular automata model
containing moveable wolves and sheep. For the implementation of
the parallel updating, we describe in detail a new definition of
neighborhood. The population is composed of individuals
characterized by their genetic strings and ages by applying the
Penna model$^{13}$ with sexual reproduction.

The paper is organized as follows: in the next section we present
the model, section 3 contains the simulation method, section 4
contains the results and discussion, and the conclusions are in
the final section.

\section{The predator-prey model with sexual reproduction}
\noindent We consider a cellular automata model composed of
wolves, sheep and grass on a square lattice of linear size $L$ ($L$ is
a common multiple of 4), with periodic boundary conditions. The
time was considered as a discrete variable ($t= 1, 2,3,\ldots $)
as suited for implementation on computers.

On a cell it is occupied by one independently moveable animal at
most, or it maybe empty, i.e., an independently moveable wolf, an
independently moveable sheep, a mother wolf or a mother sheep with
her child who can not move independently, or grass only. It is
forbidden for two or more independently moveable animals to occupy
the same cell at the same time. Grass, which grows on each cell,
can be eaten up by a sheep in one time step, and grow again after
certain time steps. The grass in our model is not always
available, which has not been considered in previous papers.

Each individual has a reserve of food, in our model it is
represented by a counter containing its "food rations". The
counter is increased to the maximum value after each "meal"
(eating grass by a sheep or eating sheep by a wolf) and decreased
by one after completing one simulation step. The animal dies when
its counter reads zero.

Another feature characterizing an individual is its genotype that
is a double bit string (of 32 positions) of 0's and 1's. It could
have the form \[  \{  \begin{array}{c}
 0010110\ldots\\0101101\ldots \end{array}
\]They are defined at birth, kept unchanged during the individual
lifetime and read in parallel.

From the genotype the phenotype of the individual is constructed,
as a single string of the same length, according to the following
rule.

All the sits are sorted as recessive and dominant sits, each of
which take up 50\% of the total sites. On the dominant sites, if a
"1" appears, i.e., (0,1), (1,0) or(1,1), a 1 is put at the
corresponding place of the phenotype, otherwise a 0. While on the
recessive sites, if a "0" appears, i.e., (0,0), (0,1) or(1,0), a 0
is put at the corresponding place of the phenotype, otherwise a 1.
For details see Table 1.

\begin{table}[htbp]
\tcaption{The phenotype determined by the genotype.}
\centerline{\footnotesize\smalllineskip \centering
\begin{tabular}{ c c c}\\
\hline
{Type of sites} &{Genotype} &{Phenotype}\\
\hline
{}  &(0,0)&0 \\
\raisebox{1.37ex}[0pt]{Dominant}&(0,1) (1,0) and(1,1) &1 \\
\hline
{}  &(1,1) &1 \\
\raisebox{1.37ex}[0pt]{Recessive} &(0,0) (0,1) and(1,0) &0 \\
\hline\\
\end{tabular}}
\end{table}

The bit on the $i$th site of the phenotype gives the information
if the individual will suffer from the effects of a genetic
disease in its $i$th year of life. Diseases are represented by 1.
In each time step, the total amount of diseases until the
individual's current age is compared with a threshold $T$ , when
it overloads this limit, the individual dies. Here we assumed a
predation rule that only when the number of the genetic diseases a
wolf suffered is less than that of a sheep, the predation can be
successful.

Each offspring receives its genotype constructed through
recombination and formation of two gametes from each parent
(genetic shuffling).The process may be described as follows. The
two strings of a parent's genotype are cut at a random place. The
resulting four pieces are glued across, forming two gametes. The
same is done for the second parent. From each parent one gamete is
chosen randomly and modified by $M$  random mutations, thus
forming two strings for the genotype of the offspring. Similar
ideas of the formation of the genotype of the offspring have been
considered.$^{17,18,23}$ Here we consider both deleterious and
helpful mutations.

If an animal has one neighbor of the same species with opposite
sex, the pair reproduce one offspring. In order to breed, apart
from finding a partner in its neighborhood, the animal must be
strong enough, i.e., it must have at least $r_{min}$   food
rations and at least reach the minimum reproduction age $A_r$  .
We call these reproduction terms. The sex of the newborn offspring
is randomly selected, and its food ration is equal to its
mother's.

Considering the strategy of child-care $^{22}$, we defined a
period during which if there is no empty site in the neighborhood
the child could stay with its mother and move together. In this
period, if there is an empty site in the neighborhood, the child
leaves its mother and moves into it. The child will die, if there
is still not any empty site in the neighborhood, after having
overloaded the period $C_c$  .  Thus fewer children survive
because of lack of space. This procedure takes care of the
unrealistic unlimited growth of a population found in the classic
Lotka-Volterra models and replaces naturally the phenomenological
Verhulst factor.

\section{Cellular automata simulation method}
\noindent A cellular automaton (CA) is a regular array of cells.
Each cell can be in one of a set of possible states. The CA
evolves in time in discrete steps by changing the states of all
cells simultaneously. The next state which a cell will take is
based on its previous state and the states of the neighboring
cells. The behavior of CA is specified by rules governing local
relationships, the set of states, the neighbors, and the
transition rules. They are an attempt to simplify the often
numerically intractable dynamic simulations into a set of simple
rules that are easy to compute. As an approach to the modeling of
properties of complex systems they have a great benefit in being
visually informative of the progress of dynamic events. From their
early development by von Neumann $^{13}$, a variety of biological
applications have been reported. $^{14-16}$

In our model containing moveable wolves and sheep, in order to
obey the CA laws in von Neumann neighborhood (Fig. 1(a)), it is
necessary disobey the common sense because a wolf may be
participant in several hunts, which is unreasonable.

To solve this problem, we present a new definition of
neighborhood. A sub-lattice is considered as an elementary
component of the whole lattice. Then there are two schemes for the
definition of the neighborhood as shown in Fig. 1(b), 1(c) and
1(d). In this way the sub-lattice is divided into eight neighbor
pairs. Only neighbor sites belonging to the same neighbor pair can
react, so all the neighbor pairs can be accessed in parallel.

\begin{figure}[htbp]
\scalebox{0.54}{\includegraphics{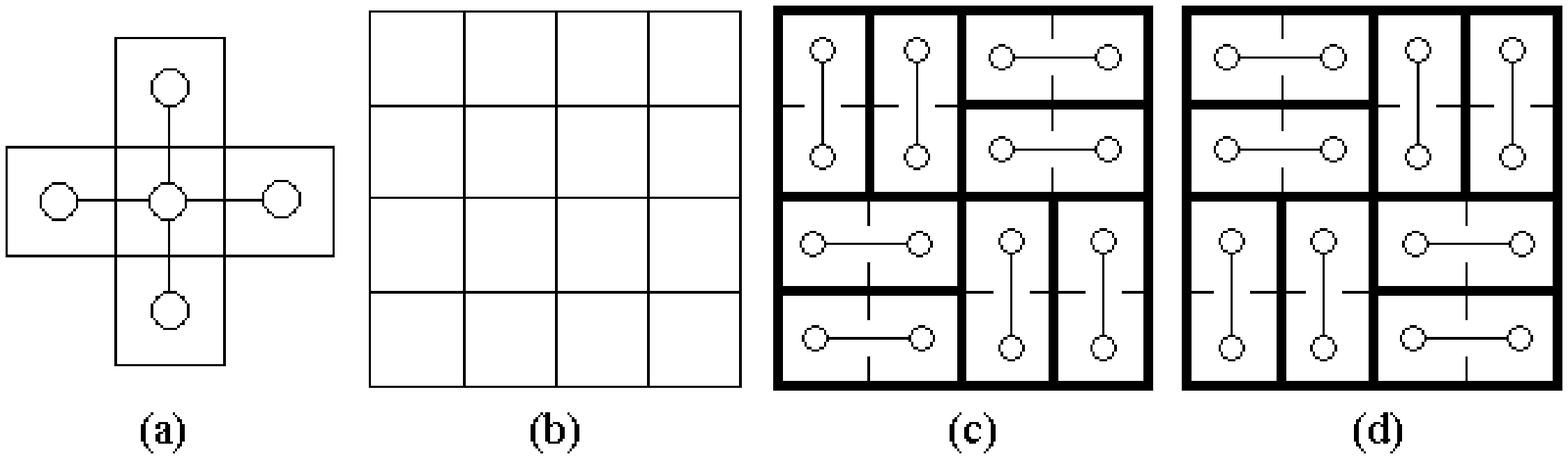}}
\fcaption{We show the traditional neighborhood and the new defined
one. (a) von Neumann neighborhood, (b) a sub-lattice is considered
as an elementary component, (c) scheme 1 for the definition of the
neighborhood, (d) scheme 2 for the definition of the
neighborhood.}
\end{figure}

Applying this rule to the  ($L$ is a common multiple of 4)
lattice, we build up a tiled mask over the whole lattice,
considering periodic boundary conditions. For any lattice, which
size is (${L=4k}$), it can be proved that there are only sixteen
possible partitions. Take a lattice for instance( Fig. 2), cells
of the same gray scale belong to the same elementary component.

\begin{figure}[htbp]
\centering \scalebox{0.7}{\includegraphics{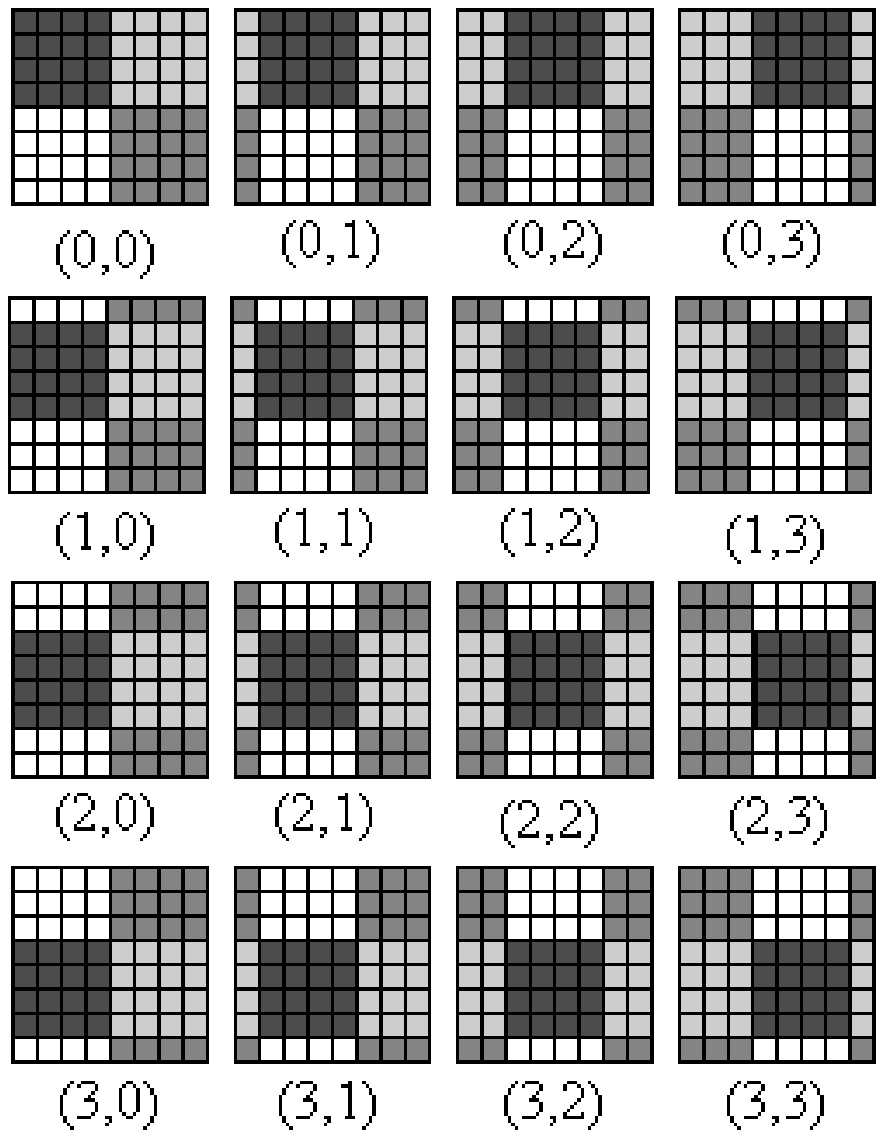}}
\fcaption{The sixteen possible partitions of elementary components
in a   lattice. Cells of the same gray scale belong to one
elementary component.}
\end{figure}

The dynamics is not confined to big blocks (elementary
components), because the boundary between blocks is changing from
one time step to the next. In this way, the implementation of this
model is more tractable than the asynchronous Cellular Automata
and more efficient than Monte Carlo simulation.

  The evolution runs along the following steps:
1. Choose randomly one of the sixteen possible partitions as shown
in Fig. 2. Inside each elementary component specify the neighbor
pairs by selecting one of the two schemes randomly. 2. Sweep over
all the neighbor pairs and inside each neighbor pair do the
reaction. 3. Increase the time by one time step. 4.   Return to
step 1. We applied this simulation method to this predator-prey
model. The states of the sites are denoted as a set of symbols,
(Table 2), and the reaction rules inside the neighbor pairs are
shown in Table 3.

\begin{table}[htbp]
\tcaption{The Symbol set according to the resident in the sites.}
\centering \centerline\footnotesize\smalllineskip
{\begin{tabular}{l c}\\
\hline
{The resident in the sites } &{Genotype} \\
\hline
{A wolf }  &$W$ \\
\hline
{A male wolf}&${W_m}$ \\
\hline
{A female wolf}&${W_f}$ \\
\hline
{A female wolf with her child under the care}&${{W_f}^*}$ \\
\hline
{A sheep}&$S$ \\
\hline
{A male sheep}&$S_m$ \\
\hline
{A female sheep}&$S_f$ \\
\hline
{A mother sheep with her child under the care}&${{S_f}_*}$ \\
\hline
{Grass only(empty site)}&$G$ \\
\hline
\end{tabular}}
\end{table}

\begin{table}[htbp]
\tcaption{Reaction rules inside the neighor pairs.($R_p$: both of
the male and female animals satisfy the reproduction terms. $S_p$:
the number of the genetic diseases the wolf suffered is less than
that of the sheep) } {
\centering
\begin{tabular}{ c c l}\\
\hline
 {Current states of}&{States of a neighbor}&{}\\
{a neighbor pair} &{pair in the next time step} &\raisebox{1.37ex}[0pt]{Interpretation}\\

 \hline
{}& {}&A sheep move into  \\
\raisebox{1.37ex}[0pt]{$S+G$} &\raisebox{1.37ex}[0pt]{$G+S$}&{ an empty site.}\\
 \hline
 {}& {}&A baby sheep leave its mother  \\
\raisebox{1.37ex}[0pt]{${{S_f}^*+G}$}&\raisebox{1.37ex}[0pt]{$S_f+S$}&{and move into an empty site.} \\
\hline
{}&{}&A wolf moves into\\
\raisebox{1.37ex}[0pt]{$W+G$}&\raisebox{1.37ex}[0pt]{$G+W$}&{ an empty site. }\\
 \hline
{}&{}&A baby wolf leaves its mother\\
\raisebox{1.37ex}[0pt]{$W_f^*+G$}&\raisebox{1.37ex}[0pt]{$W_f+W$}&{and moves into an empty site.}\\
\hline
{}&{}&A pair of male and female sheep \\
\raisebox{0ex}[0pt]{$S_m+S_f$}&\raisebox{0ex}[0pt]{$S_m+S_f^*$} &{reproduce an offspring }\\
{}&{}&{when they satisfy the $R_p$ term.}\\
 \hline
{}&{}&A wolf eats up a sheep\\
\raisebox{1.37ex}[0pt]{$W+S$}&\raisebox{1.37ex}[0pt]{$W+G$}&{when they satisfy the $S_p$ term.}\\
\hline
{}&{}&A mother wolf eats up a sheep,\\
\raisebox{1.37ex}[0pt]{$S+W_f^*$}&\raisebox{1.37ex}[0pt]{$W+W_f$}&{then the baby wolf left mother.}\\
\hline
{}&{}& A pair of male and female \\
\raisebox{0ex}[0pt]{$W_m+W_f$}&\raisebox{0ex}[0pt]{$Wm+W_f^*$}&{
wolves reproduce an offspring }\\
{}&{}&{when they satisfy the $R_p$ term.}\\
 \hline
\raisebox{0ex}[0pt]{$G+G$}&{} &{}\\
\
\raisebox{0ex}[0pt]{$S_m+S_m$}&{} &{} \\
\
\raisebox{0ex}[0pt]{$S_f+S_f$}&{} &{} \\
\
\raisebox{0ex}[0pt]{$S_f^*+S_f^* $}&{} &{} \\
\
\raisebox{0ex}[0pt]{$W_m+W_m$}&{}&{Under these cases,} \\
\
\raisebox{0ex}[0pt]{$W_f+W_f $}&{}&{ keep unchanged.} \\
\
\raisebox{0ex}[0pt]{$W_f^*+W_f^*$}&{}&{} \\
\
\raisebox{0ex}[0pt]{$S_m+S_f^*$}&{}&{} \\
\
\raisebox{0ex}[0pt]{$S_f+S_f^*$}&{}&{} \\
\
\raisebox{0.3ex}[0pt]{$W_m+W_f^*$}&{}&{} \\
\
\raisebox{0.3ex}[0pt]{$W_f+W_f^*$}&\raisebox{15ex}[0pt]{---}&{} \\
\hline\\
\end{tabular}}
\end{table}

The parameters of the model are as follows: linear size of the
lattice L, which is a common multiple of 4; initial densities of
both species: prey (sheep) ${C_s(0)}$  and predator (wolves)
$C_w(0)$; the number of time steps  needed for the grass to grow
again $g$; the maximum food ration$k$; the amount of food $r_min$
needed to be fit for breeding; the minimum reproduction age $A_r$;
the mutation rate $M$; the limit number of genetic disease $T$;
the period $C_c$ for the new born animal under child-care. For
simplicity, we assume the same value of  $k$, $r_min$, $A_r$, $M$,
$T$ and $C_c$ and  for predators and prey. To make the model
tractable, we fixed some of the parameters at the following
reasonable values:$g=4$,$k=3$,$r_min=2$, $A_r=8$, $T=3$,$C_c=3$.

The initial populations are random, i.e. , they have random
genotypes, ages and food rations. The spatial distribution is
random, too, with the given initial concentrations.

\section{Results and discussion}
\noindent As could be expected, we have found three possible
states: the coexisting one with prey and predators, the absorbing
one with prey only, and the empty one where no animal survived,
Fig. 3(a), 3(b) and 3(c), The fate of the two species mainly
depends on their initial densities.
  The first typical stationary state is shown in Fig. 3(a)
  (which is the most usual state for the system to achieve in the model),
   which the oscillatory coexistence of the system is shown out.
   The three curves represent the densities of the predators,
   the prey and the grass respectively. The oscillations about the densities of the wolves,
    the sheep and the grass are not damped away and persist with big  amplitude.

The second typical stationary state is shown in Fig. 3(b). The
wolves die out rapidly, leaving the sheep and the grass. There are
many factors resulting in the extinction of the wolves. The
critical one is that the initial density of the sheep is too low
while the one of the wolf is too high. After capturing most of the
sheep, the wolves starve to death because of lack of food. The
survived sheep spread rapidly after the wolf's extinction. In this
case the sheep and the grass reach an asymptotical stationary
state. Compared with Penna models without lattice $^{5,13}$, the
oscillation of the density of sheep in this model are not damped
away even after 2000 time steps.

  The third typical stationary state is shown in Fig. 3(c).
  The initial density of the wolves is so high that they
   eat up all the sheep and subsequently starve to death.
    Finally it leads to the extinction of both of the species.

  Then we investigated the effect of the mutation on the evolution of the two species.
   In Fig. 4(a), 4(b) and 4(c), $C_s(0)=0.3$, $C_w(0)=0.3$, $L=256$,
    $M=1, 2\; and \;4$ respectively. $N_{bg}$ is the average number of "1"
    in the phenotype of the species. One can see that the $N_{bg}$ drops with time step.
     What is the reason? There are mainly two factors. First,
     the mutation ensures the genetic diversity, including deleterious and helpful mutations.
     And then, the predation rule we assumed in this model plays the role of natural selection,
     that is, only the animals with better genotype are able to survive and reproduce.
     From the comparison of the three figures, one can see increasing
     the mutation rate leads to the faster evolution.

Now we discuss the role played by the linear size of the lattice
on the type of the steady state reached by the system. The
possible range of the initial concentrations is defined as
$A={(x_1,x_2)|(x_1+x_2<1,0<x_1<1,0<x_2<1)}$. The interval from 0
to 1 can be divided evenly into $N$ points. The position of each
point represents the possible initial densities of the wolves and
the sheep, and its color represents the steady state led by the
initial densities.
  The $N*N/2$ points are sorted according to the final states
  they reach after 3000 simulation steps
  (Fig. 5(a), 5(b) and 5(c)) (2000 simulation steps is enough by plenty of experiments).
   From the three figures with different value of $L$, we concluded that the bigger
   the linear size of the lattice $L$ is, the larger the proportion of the coexistence area will be.

Snapshots are given to show the spatial evolution of this
predator-prey system, Fig. 6(a), 6(b) and 6(c). The initial random
distribution of the wolves and sheep ($C_s=0.3,C_w=0.3,M=2,L=256$)
is shown in Fig. 6(a). After 100 and 200 time steps, spatial
self-organization can be observed in Fig. 6(b) and Fig. 6(c)

\section{Conclusions}
\noindent We have presented a lattice model of prey and predators
that characterized by Penna bit-string. To survive, each animal
has to eat at least certain time steps. A pair of male and female
animals in adjacent cells can reproduce an offspring, if they have
enough food reserves and over the minimum reproduction age. The
child-care strategy allows the offspring to stay and move together
with its mother within a certain period, if there is no empty cell
in the neighborhood. Most of the predator-prey lattice models are
simulated via Monte Carlo method. In our model containing moveable
wolves and sheep, in order to obey the CA laws in von Neumann
neighborhood, it is necessary disobey the common sense because a
wolf may be participant in several hunts, which is unreasonable.
To solve this problem, we present a new definition of neighborhood
and built a fully parallel cellular automata simulation method,
which is more tractable than the asynchronous Cellular Automata
and more efficient than Monte Carlo simulation. As could be
expected, we have found three possible states: the coexisting one
with prey and predators, the absorbing one with prey only, and the
empty one where no animal survive. In addition, we discussed the
final states attained by the population with certain initial
densities, the effect of the mutation on the evolution of the two
species and the role played by the linear size of the lattice on
the type of the steady state reached by the system.

\nonumsection{References} 

\begin{figure}[htbp]
\vspace*{10pt}\scalebox{0.5}{\includegraphics{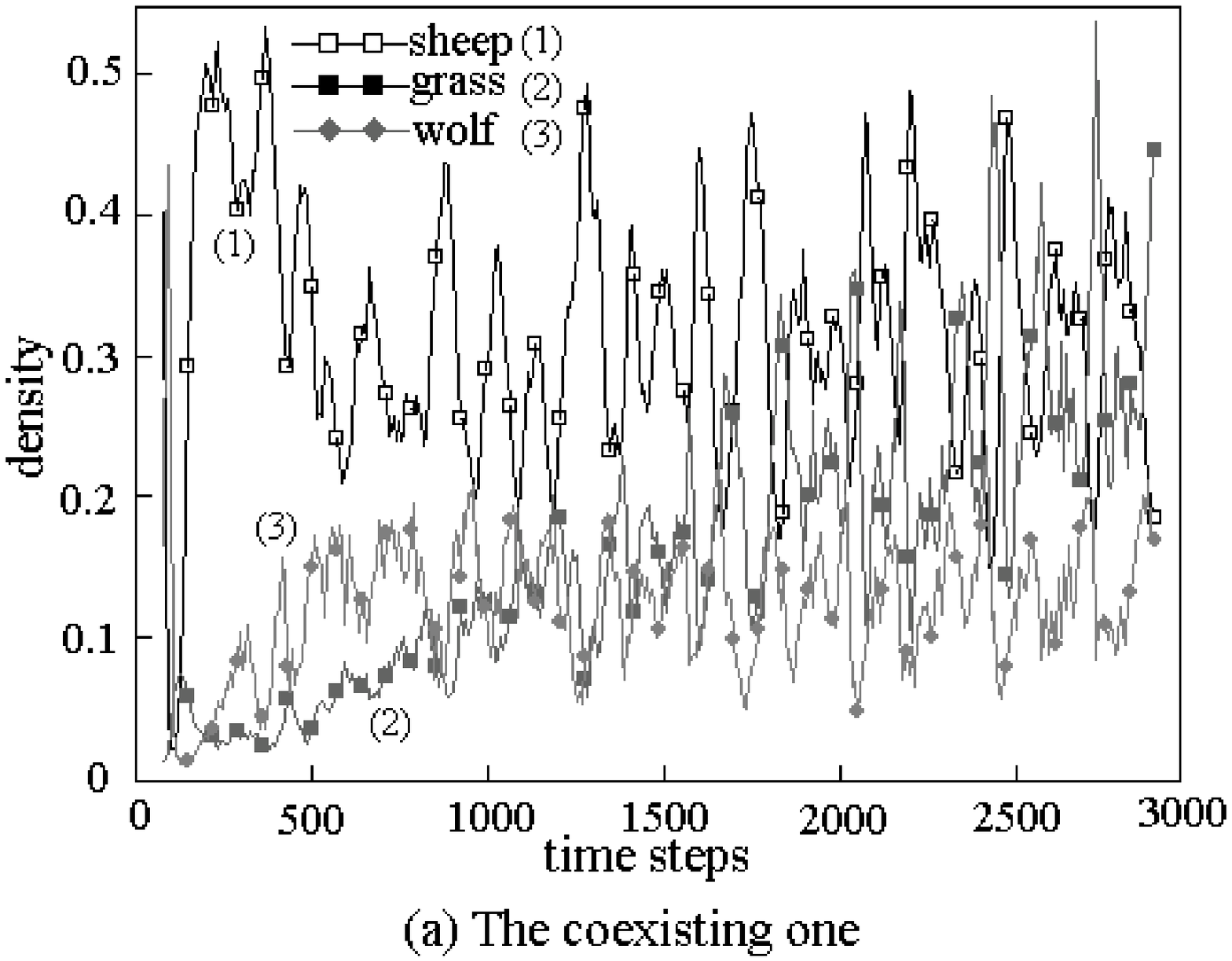}}
 \centering
\vspace*{10pt}\scalebox{0.5}{\includegraphics{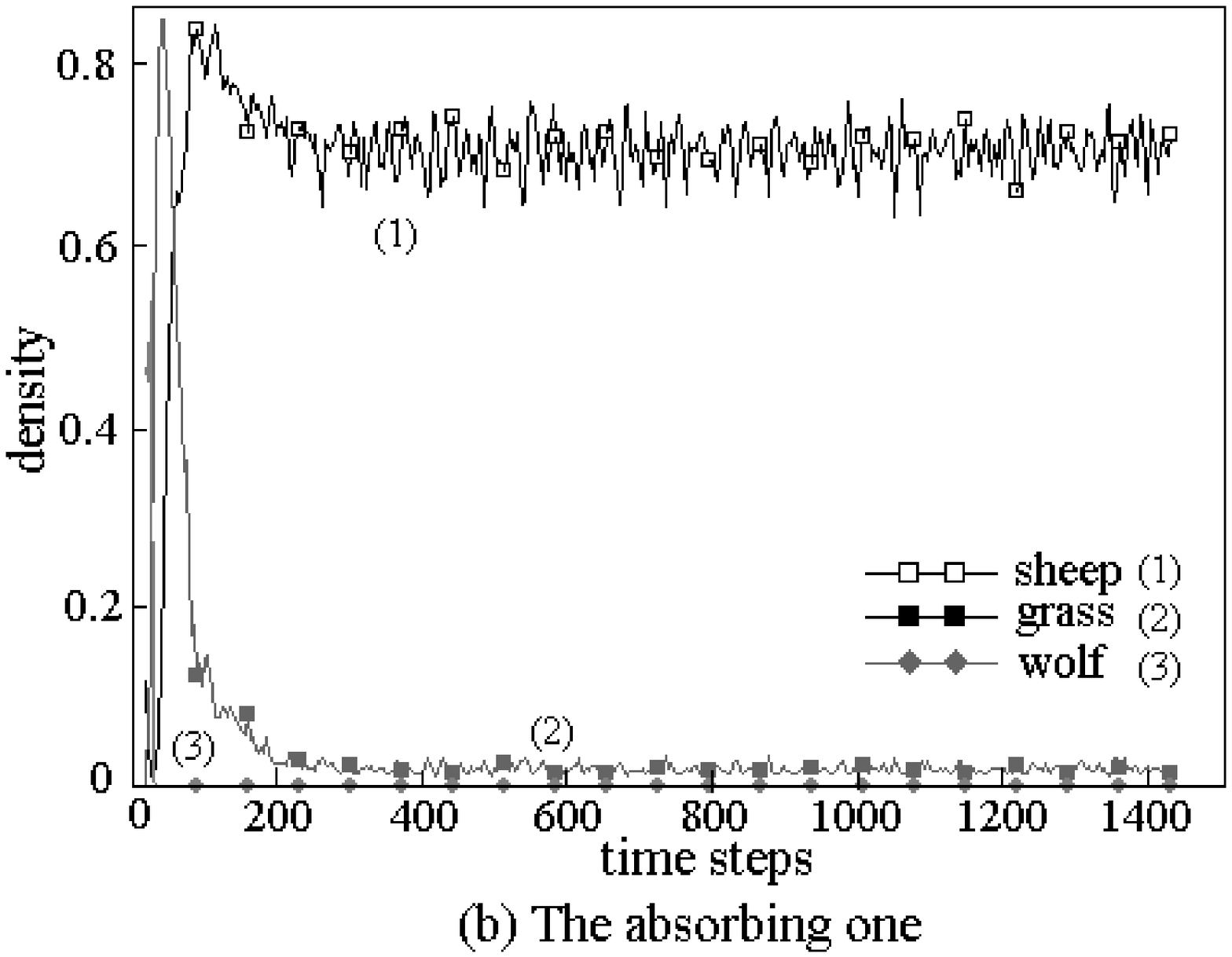}}
\centering
\vspace*{10pt}\scalebox{0.5}{\includegraphics{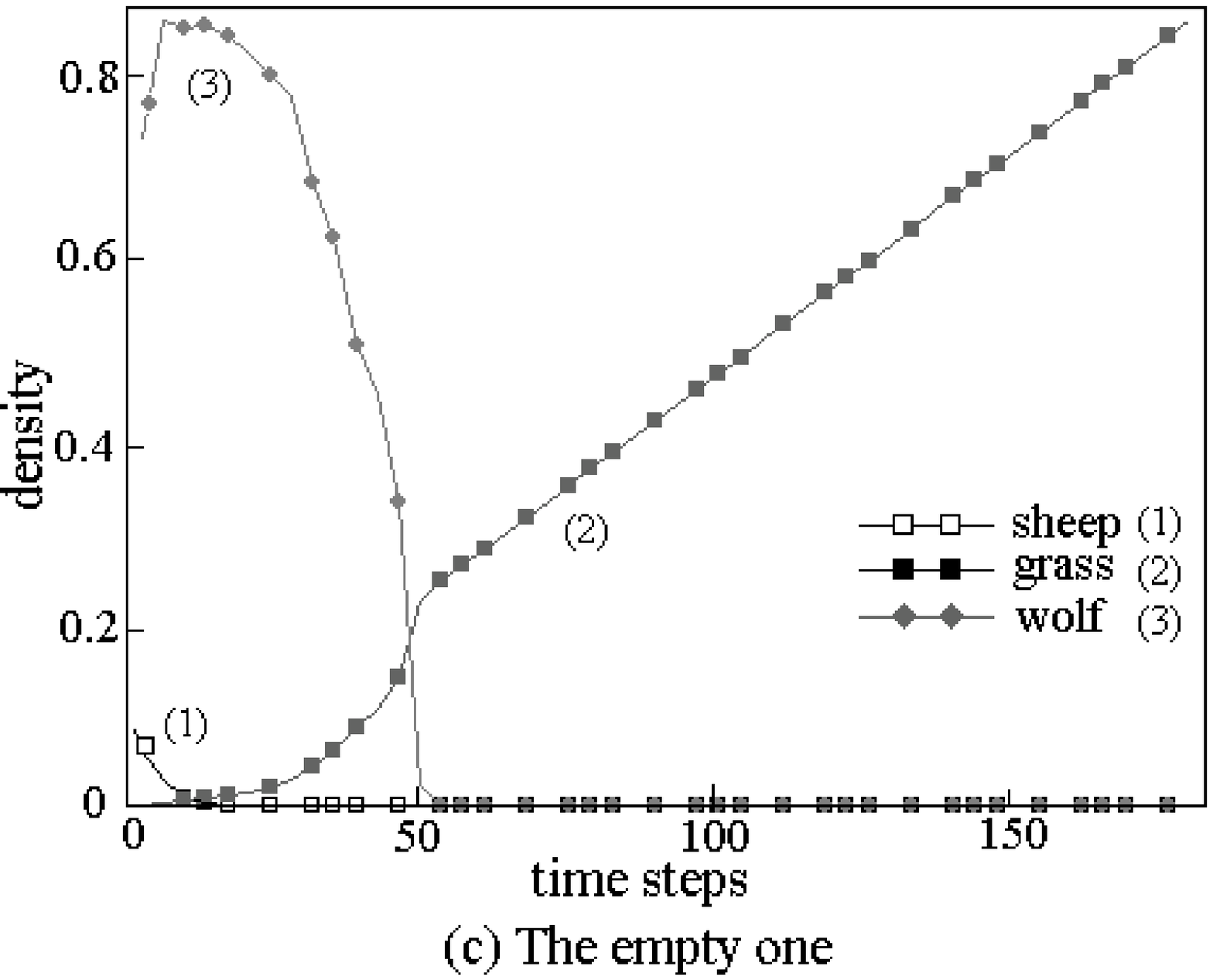}}
\fcaption{\footnotesize Density versus time step,$M=2$,$L=256$}
\centerline{\footnotesize (a)the final state is the prey only one,
$C_s(0)=0.3$, $C_w(0)=0.3$;} \centerline{\footnotesize(b)the final
state is the empty one.$C_s(0)=0.1$, $C_w(0)=0.8$ ;}
\centerline{\footnotesize(c)the final state is the empty one,
$C_s(0)=0.1$, $C_w(0)=0.8$.}
\end{figure}

\begin{figure}[htbp]
\scalebox{0.3}{\includegraphics{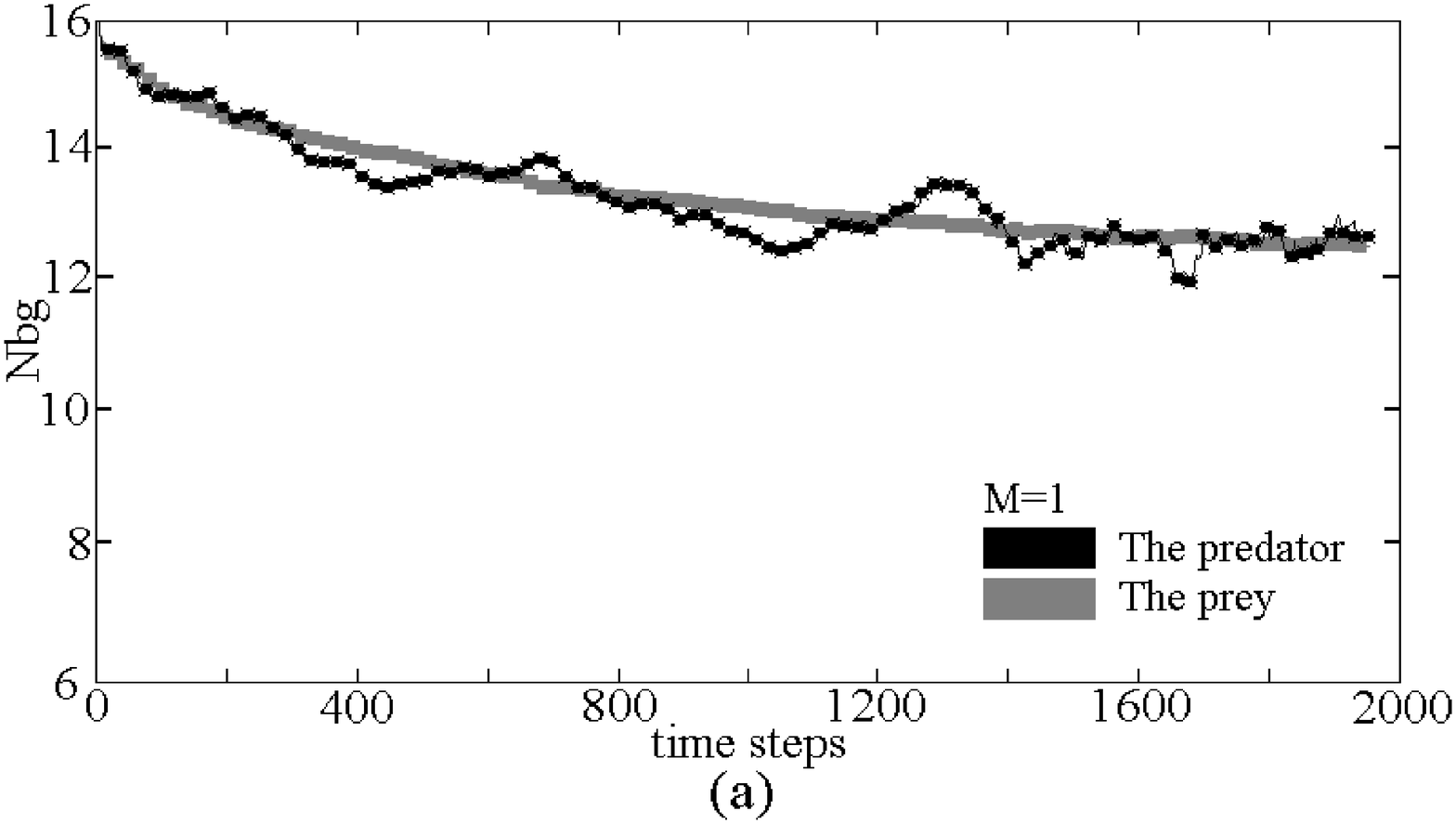}}
 \centering
\scalebox{0.3}{\includegraphics{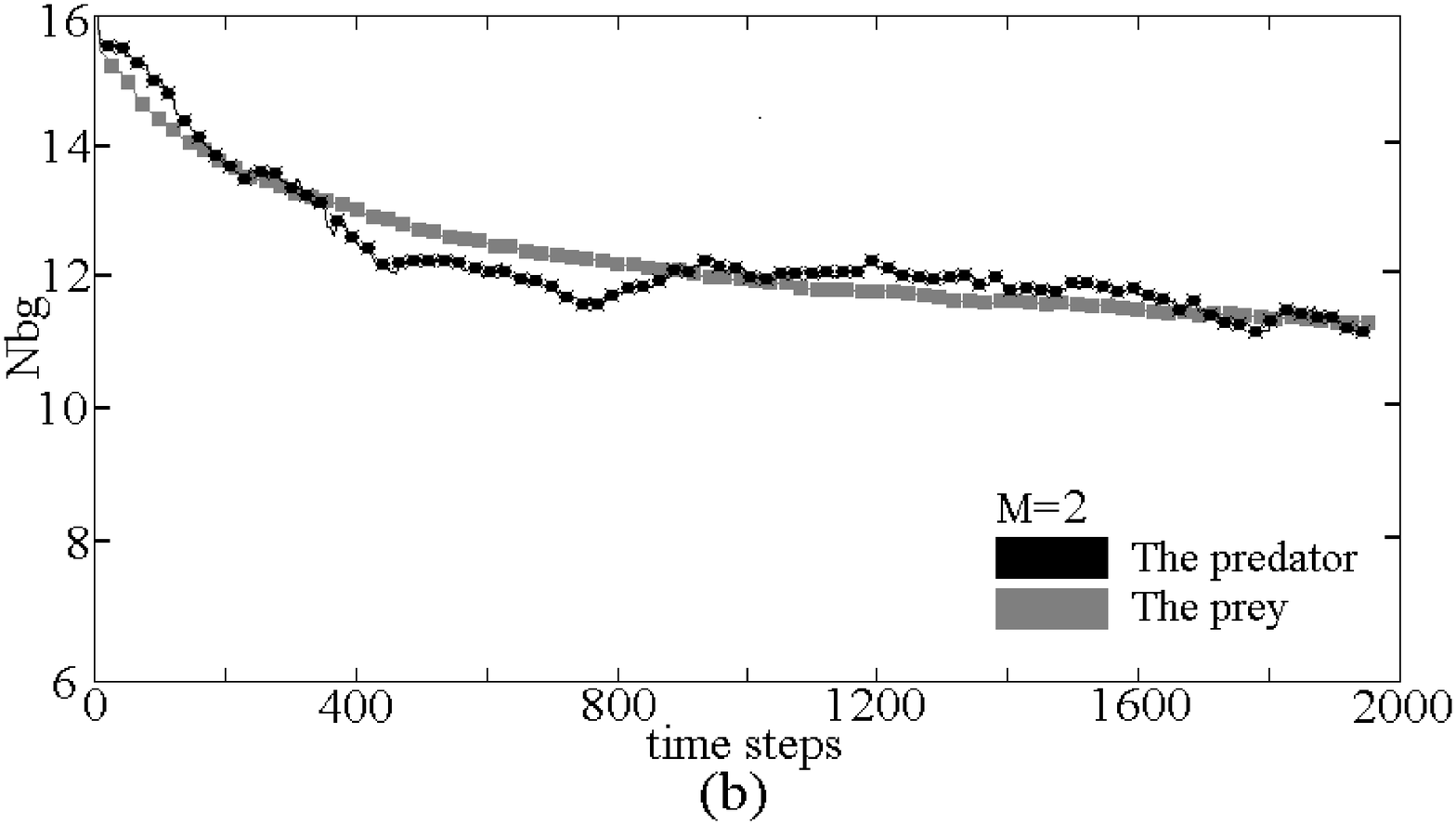}} \centering
\scalebox{0.3}{\includegraphics{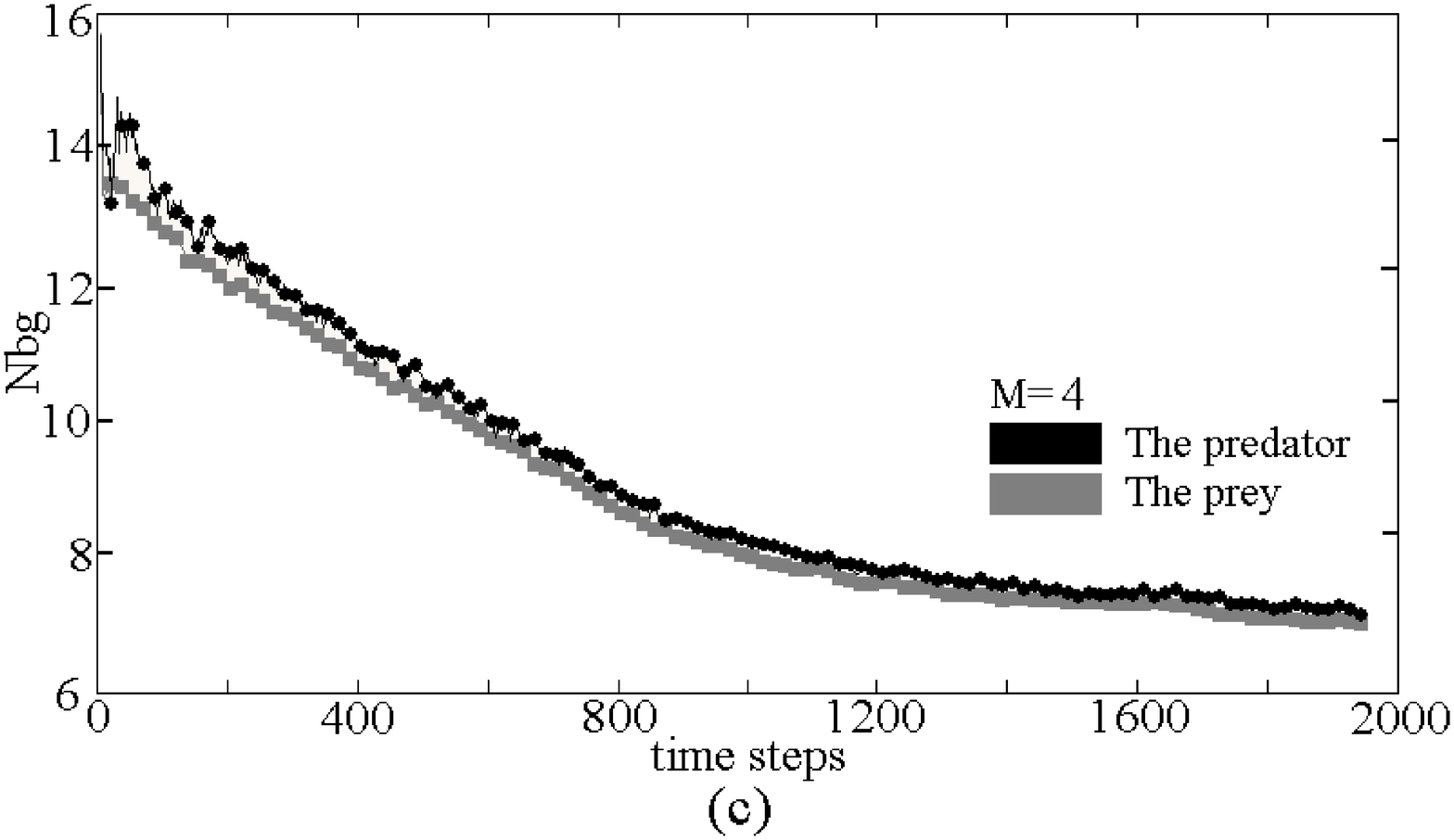}}
\fcaption{\footnotesize Time dependence of the $N_{bg}$ of the
predators and prey,} \centerline{\footnotesize $C_w(0)=0.3$,
$C_s(0)=0.3, L=256 $, (a) $M=1$, (b) $M=2$, (c) $M=4$. }
\end{figure}

\begin{figure}[htbp]
\vspace*{5pt}\centering
\scalebox{0.2}{\includegraphics{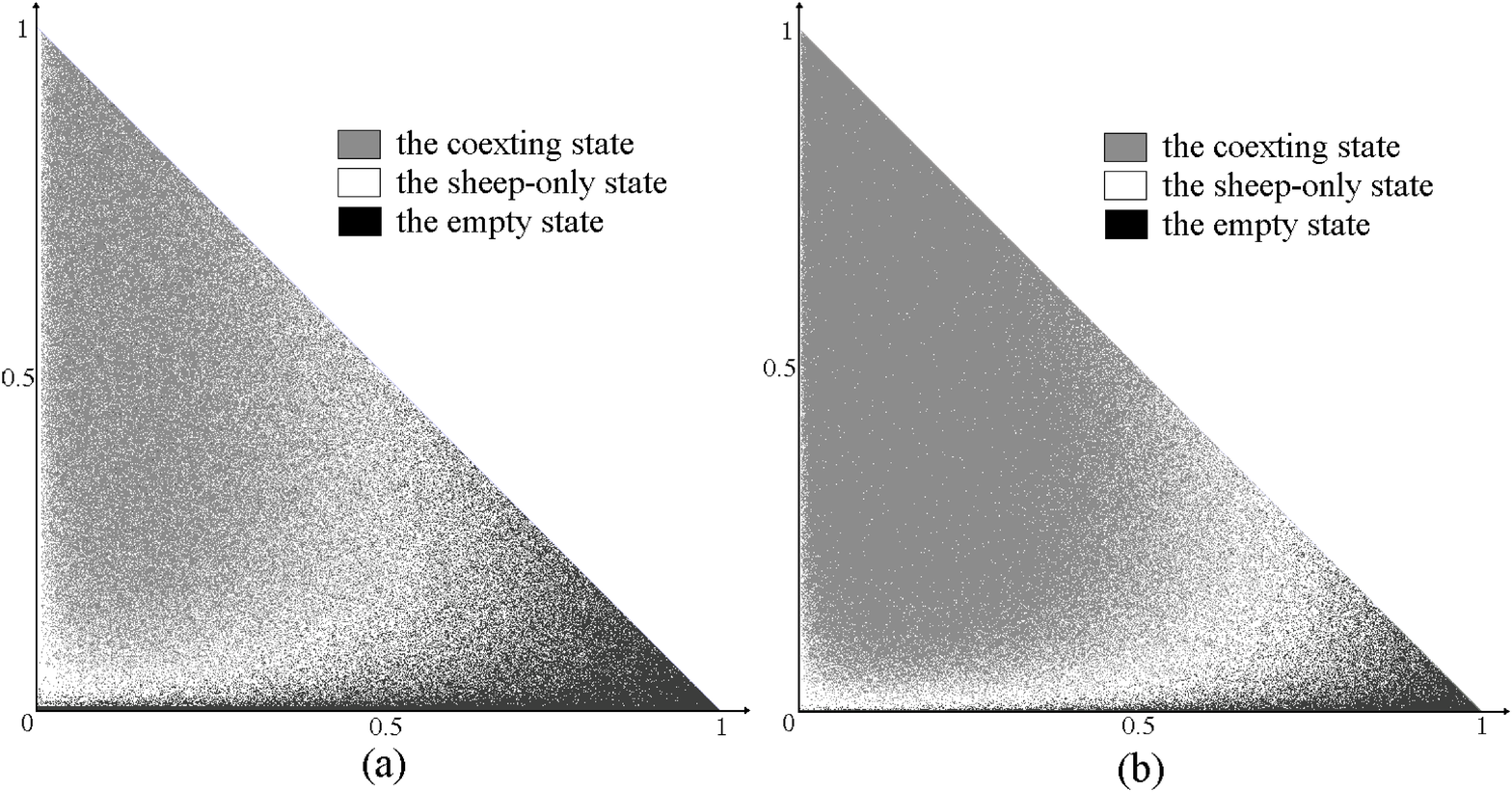}}
 \centering
\scalebox{0.2}{\includegraphics{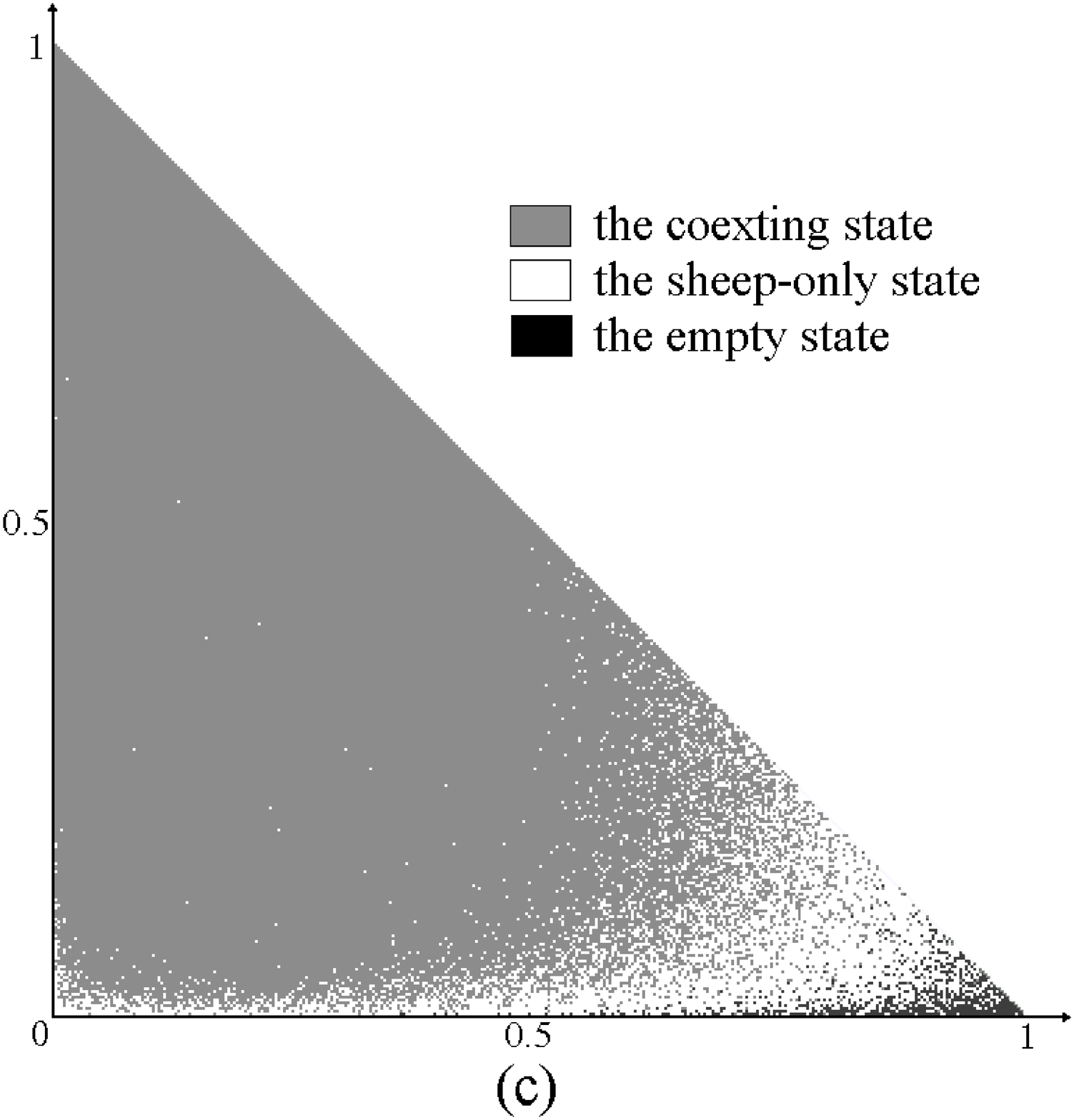}} \centering

\fcaption{\footnotesize (a) $L=128,\;M=2$ . The proportion of
black, gray and white respectively is 22.3\%, 41.8\%, 35.9\%.}
\centerline{\footnotesize (b) $L=256,\;M=2$ .The proportion of
 black, gray and white respectively is 22.3\%, 41.8\%, 35.9\%.}
\centerline{\footnotesize (c) $L=512,\;M=2$ . The proportion of
black, gray and white respectively is 2.8\%, 81.8\% , 15.4\%.}
\end{figure}

\begin{figure}[htbp]
\vspace*{5pt}\centering
\scalebox{0.5}{\includegraphics{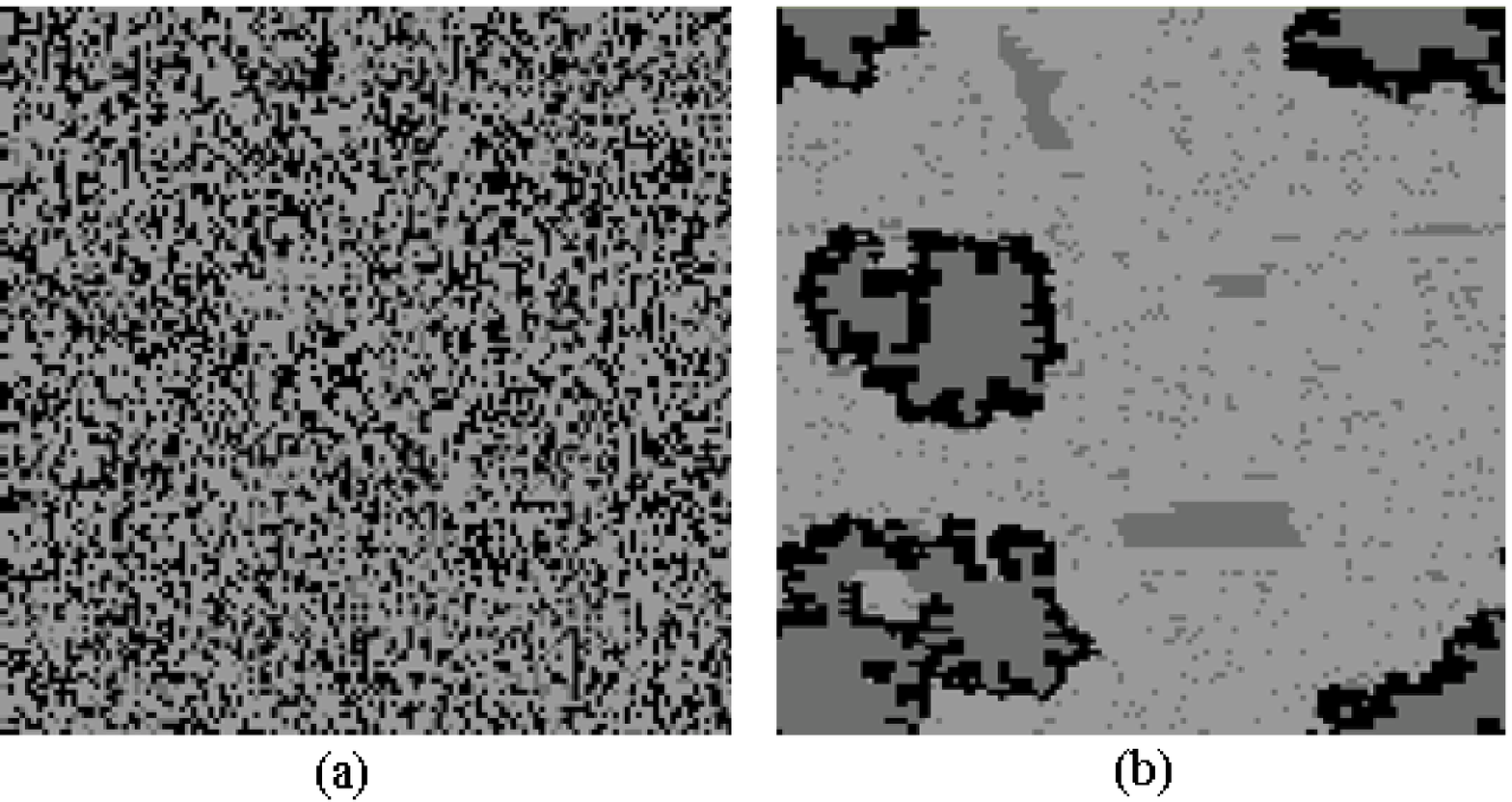}}
 \centering
\scalebox{0.5}{\includegraphics{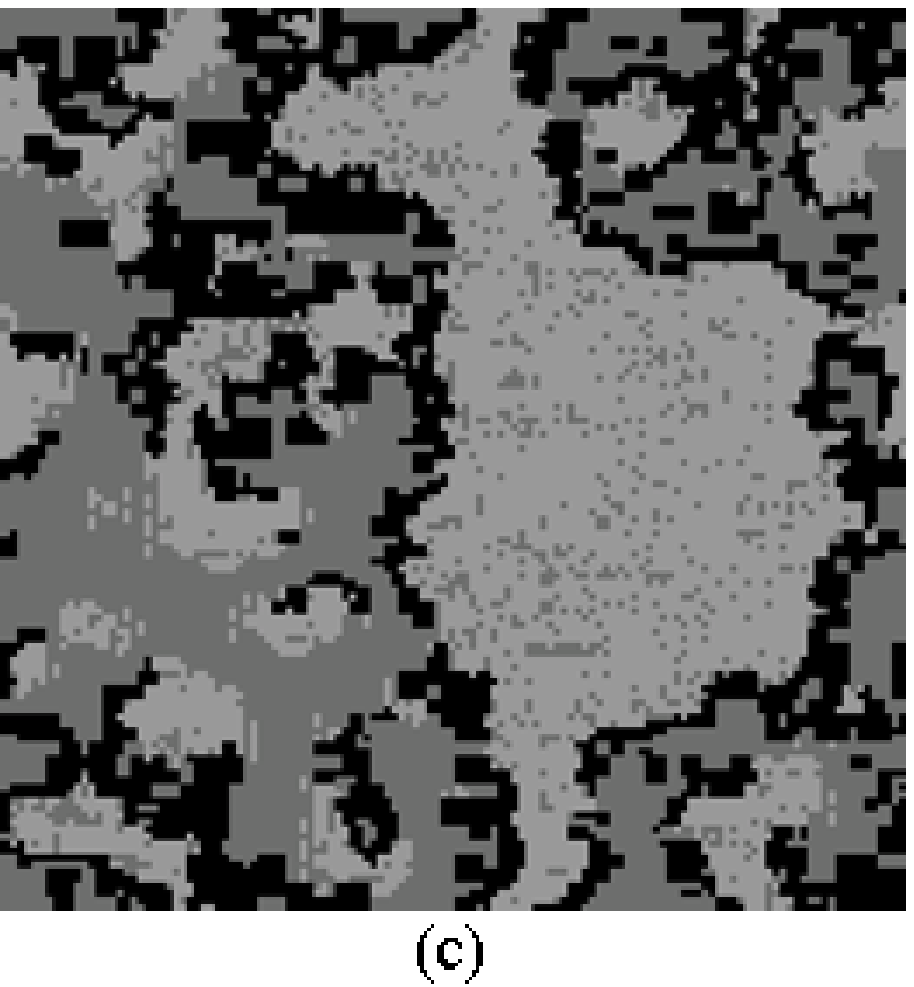}} \centering
\fcaption{\footnotesize (a) At the initial state the wolves and
the sheep are distributed randomly. }
 \centerline{\footnotesize (b) After 100 time steps,self-organization can be observed.}
\centerline{\footnotesize (c)The distribution of the grass, sheep
and wolf after 200 time steps.}
\end{figure}

\end{document}